\documentclass[aps,prl,amsmath, amssymb, english, aps, prl, onecolumn,
  reprint,floatfix, superscriptaddress]{revtex4-1}

\usepackage[english]{babel}
\usepackage{graphicx}
\usepackage{verbatim}

\usepackage{epsfig}
\usepackage{epstopdf}
\usepackage{amsfonts}
\usepackage{amsthm}
\usepackage{amsmath}
\usepackage{amssymb}
\usepackage{hyperref} 
\hypersetup{colorlinks=true}
\usepackage{makeidx}
\usepackage{pifont}

\usepackage{subfig}

\usepackage{color}

\usepackage[subsectionbib]{bibunits}


\def\t2g{t${}_{2g}$}
\def\eg{e$_{g}\,$}

\begin{document}

\title{Janus-faced influence of the Hund's rule coupling in strongly correlated materials.}


\author{Luca de' Medici}
\address{Laboratoire de Physique des Solides, UMR8502 CNRS-Universit\'e Paris-Sud, Orsay, France}
\author{Jernej Mravlje}
\address{Centre de Physique Th\'eorique, \'Ecole Polytechnique, CNRS,
  91128 Palaiseau Cedex, France}
\address{Jo\v{z}ef Stefan Institute, Jamova~39, SI-1000, Ljubljana, Slovenia}
\author{Antoine Georges}
\address{Centre de Physique Th\'eorique, \'Ecole Polytechnique, CNRS, 91128 Palaiseau Cedex, France}
\address{Coll\`ege de France, 11 place Marcelin Berthelot, 75005 Paris, France}
\address{Japan Science and Technology Agency, CREST, Kawaguchi
  332-0012, Japan}

\begin{abstract}
We show that in multi-band metals the correlations are strongly
affected by the Hund's rule coupling, which depending on the filling
promotes metallic, insulating or bad-metallic behavior. The
quasiparticle coherence and the proximity to a Mott insulator are
influenced distinctly and, away from single- and half-filling, in
opposite ways. A strongly correlated bad-metal far from a Mott phase
is found there.
We propose a concise classification of 3d and 4d                                                         
transition-metal oxides within which the ubiquitous                                                    
occurrence of strong correlations in $\mbox{Ru}$- and                                                                    
$\mbox{Cr}$-based oxides, as well as the recently measured high N\'eel                                                         
temperatures in $\mbox{Tc}$-based perovskites are naturally explained.
\end{abstract}

\maketitle

 Hund's rules determine the ground-state of an isolated atom by
 accounting for the dependence of the Coulomb repulsion between
 electrons on their relative spin and orbital configurations.  In
 insulating solids, their role is to select the relevant atomic
 multiplets, which are then coupled by inter-site magnetic
 interactions.  In contrast, the effects of the Hund's rule coupling in
 metallic compounds are less understood.  The difficulty lies in
 dealing with the localized (atomic) and itinerant characters of
 electrons on equal footing, a key issue for materials with strong
 electron correlations~\cite{imada_mit_review}.  Despite increasing
 awareness of the physical relevance of the Hund's rule coupling for such
 materials\cite{haule09,werner08,werner09,mazin_Hund_pnictides,aichhorn10,mravlje11,demedici_MottHund},
 a global view is still lacking. 

 In this article, we fill this gap
 and provide a classification with respect to the number of electrons
 filling the active orbitals.  We show that, aside from the case of a
 singly-occupied shell (where metallicity is
 favored\cite{werner09,demedici_MottHund}), or a half-filled shell
 (where it promotes Mott insulating
 behaviour\cite{bunemann_gutzwiller_prb_1998,han_multiorbital}) the
 Hund's rule coupling induces conflicting tendencies and thus causes strong
 correlations far from the insulating state (`bad-metal'
 behavior).  This picture explains the observed physical properties of
 a number of transition-metal oxides and allows for predictions on
 novel ones, such as Technetium compounds.

In order to describe all these possibilities and
illustrate our key-point in a simple context, we 
consider a
model of three identical bands with semicircular density-of-states of
half-bandwidth $D$ filled by $N$ electrons per site.  This is
relevant, for example, to transition-metal oxides with cubic symmetry
and a partially filled \t2g shell well separated from the empty \eg
shell.  The standard interaction Hamiltonian
\cite{imada_mit_review} can be written as:
\begin{equation}
  H_{\rm int}= (U-3J) \frac{\hat{N}(\hat{N}-1)}{2}  
   -2J \vec{S}^2 -\frac{1}{2}J \vec{T}^2 
\label{eq:hami}
\end{equation}
where $\hat{N}$ denotes total charge, $\vec{S}$
spin and $\vec{T}$ the angular momentum operators.  $U$ is the
intra-orbital interaction and $J$ is the Hund's rule coupling.  The Hund's rule 
coupling favors, in decreasing order: configurations with parallel
spins in different orbitals, with parallel spins in the same orbital,
and with opposite spins in the same orbital, maximizing $S$ and then
$T$.  We solve the model using dynamical mean-field theory
(DMFT)\cite{georges96} which maps a correlated electron system onto a quantum-impurity
problem: an effective atom coupled to a self-consistent
environment.
This approach handles the on-site atomic physics and
the itinerant motion of electrons on equal footing.
 Both the limit of an isolated atom with its multiplet
structure and that of a non-interacting band are correctly reproduced.

\begin{figure*}
{\centering
\subfloat[$N=1$\hspace{0.27\textwidth}(b) $N=2$\hspace{0.27\textwidth}(c) $N=3$]{\includegraphics[width=\textwidth]{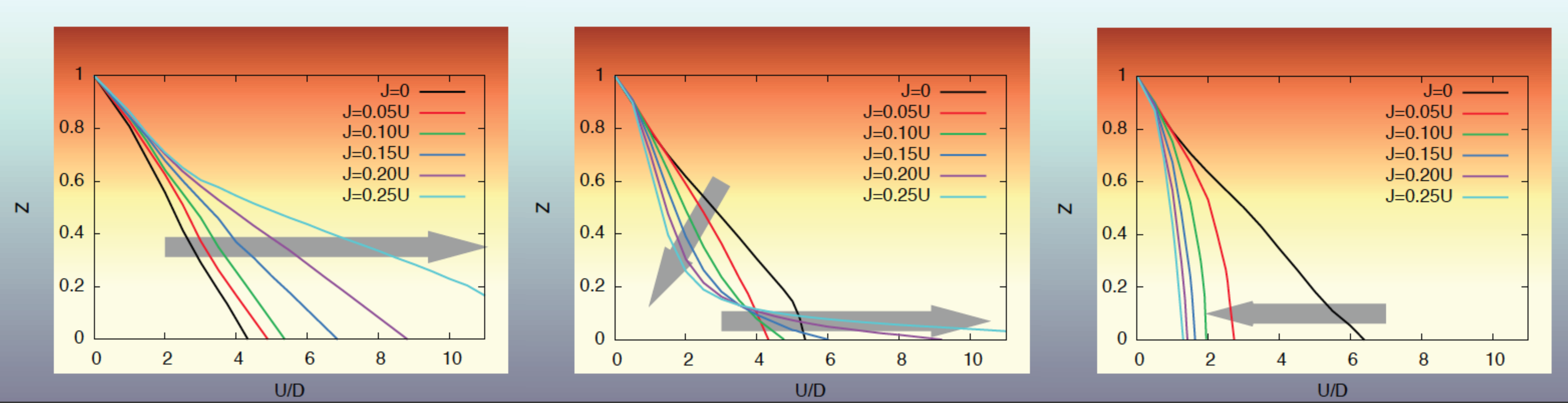}}}
  \caption{Quasiparticle weight $Z$ {\it vs.} $U$ for $N=1,2,3$ electrons in $M=3$ orbitals. The grey arrows indicate the influence of an increasing Hund's rule coupling J/U.}
  \label{fig:Z_vs_U}
\end{figure*}

In Fig.~\ref{fig:Z_vs_U}, we display the quasiparticle spectral weight
$Z$ as a function of coupling $U$, for several values of $J$, in the
paramagnetic state.  $Z$ characterizes the degree of correlations of
the metallic state. For example, the Drude weight measured from
optical conductivity is proportional to $Z$
\cite{georges96}. It also sets the energy/temperature scale $T^*$
above which the lifetime of quasiparticles becomes short and coherence
is lost.  $Z$ and $T^*$ diminish progressively as $U/D$ is increased
and a vanishing $Z$ signals the Mott insulating state, reached for
$U>U_c$.
We observe that the case of one-electron ($N=1$) is in striking
contrast to the case of a half-filled shell ($N=M=3$): upon increasing
$J/U$ the critical coupling $U_c$ in the former is
increased\cite{demedici_MottHund} while in the latter it is
reduced\cite{bunemann_gutzwiller_prb_1998,han_multiorbital}.
Correspondingly, for a fixed $U$ in the metallic phase, $Z$ increases
with $J$ for $N=1$ while it decreases for $N=M$.  Hence, as summarized
in Table~\ref{table:influence_on_coherence}, correlations are
increased by the Hund's rule coupling for a half-filled shell, and
diminished for a singly-occupied one.

\begin{table*}
\begin{ruledtabular}
\begin{tabular}{c c c c c}
Number $N$ of electrons & Degeneracy & Mott gap  & Correlations  
& Materials behaviour \\ 
in $M$ orbitals & of atomic ground-state & &  
& promoted by J\\
\hline\hline
one electron or one hole ($N=1,2M-1$) & unaffected & reduced & diminished & metallic\\ \hline
half-filled ($N=M$) & reduced & increased & increased & insulating\\ \hline
All other cases  & reduced & reduced  &  Conflicting effect & bad metallic\\
($N\neq 1, M, 2M-1$) & &  & (see text) \\
\end{tabular}
\end{ruledtabular}
\caption{The effects of an increasing Hund's rule coupling on the degree of correlations. \label{table:influence_on_coherence}} 
\end{table*}

In contrast, we find that the Hund's rule coupling has a more complex
influence in the case of two electrons (or two holes) in $3$
orbitals. On the one hand, the critical coupling $U_c$ is strongly
increased (and the Mott gap reduced) at the largest values of $J/U$.
As a result, the range of coupling $U/D$ with metallic behaviour is
enlarged as compared to the case with $J=0$. On the other hand, for a
wide range of coupling strengths, $Z$ is suppressed upon increasing
$J$\cite{mravlje11,pruschke05,lombardo_Hund_prb_2005}.
To accommodate these antagonistic effects, $Z$ displays a long tail as
a function of $U$. 
The small values of $Z$ found there indicate a very low quasiparticle coherence scale  $T^*$, below which 
the system is expected to show a conventional Fermi liquid physics.
An incoherent regime with a Curie-like magnetic response\cite{werner08} and 
bad-metal behavior \cite{werner08,mravlje11} is found for temperatures above T* (see 
the Supporting online information\cite{Janus_Supp}). This was reported in Ref. \cite{werner08} and coined 'spin-freezing' 
regime.
%
These considerations are not specific to
2 electrons in 3 orbitals: the Hund's coupling is `Janus-faced' for
$N$ electrons in $M$ degenerate orbitals, except for a singly-occupied
or half-filled band, as summarized in
Table.~\ref{table:influence_on_coherence}.

\begin{figure*}
{
 \begin{center}
   \includegraphics[width=15cm]{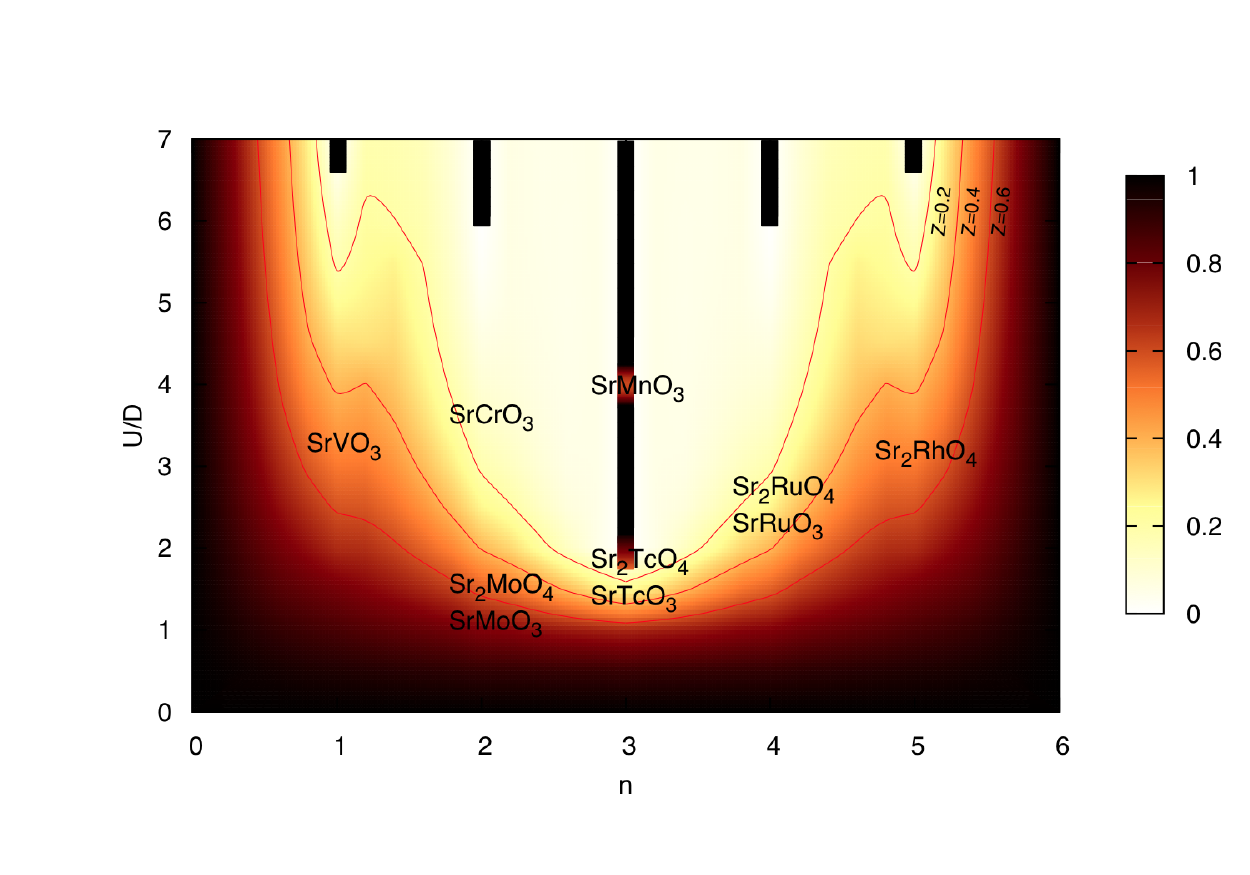}
   \end{center}
}
 \caption{Quasiparticle weight Z in a model with 3 orbitals, for
   $J/U=0.15$, as a function of the interaction strength $U$ and the
   number of electrons - from empty (0) to full (6).  Darker regions
   correspond to good metals and lighter regions to bad metals.  The
   black bars signal the Mott-insulating phases.  One notes that,
   among integer fillings, the case of 2 electrons (2 holes) displays
   bad-metal behaviour in an extended range of coupling.  Specific
   materials are schematically placed on the diagram (see text).}
\label{fig:Z_t2g_contour}

\end{figure*}

We calculated also away from integer fillings and display the data as
a contour plot of the quasiparticle weight as a function of coupling
strength $U/D$ and band filling, for a fixed typical value of the
ratio $J/U$ (Fig.~\ref{fig:Z_t2g_contour}).  The extended region of
strongly-correlated/bad metallic behaviour (small $Z$) around $N=2$
appears clearly.  In contrast, the half-filled case favors insulators
(except at weak $U/D$) and the single-electron case favors good metals
(except at strong $U/D$).

These numerical results can be corroborated and explained by analytical
considerations in the simplified limits.  In order to understand the
influence of $J$ on the Mott gap, we start from the limit of an
isolated atom.  The charge gap $\Delta_\mathrm{at}=
E_\mathrm{at}(N+1)+E_\mathrm{at}(N-1)- 2 E_\mathrm{at}(N)$ takes two
different values depending on whether the relevant orbitals are half
filled or not; $\Delta_\mathrm{at}= U-3J$ for $N<M$ or $N>M$ and
$\Delta_\mathrm{at}= U+(M-1) J$ for $N=M$.  Including hopping
perturbatively leads to a correction
$\Delta_\mathrm{Mott}=\Delta_\mathrm{at}-cD+\cdots$, where $c$ is of
order unity.
Hence, we see that the Mott gap is increased by $J$ (and $U_c$
decreased) at
half-filling\cite{bunemann_gutzwiller_prb_1998,han_multiorbital},
while it is decreased in all other
cases\cite{werner09,demedici_MottHund}.  This localized limit explains
the distinction between $N=3$ and $N=1$ (and the insulating side of
$N=2$) but does not account for the bad-metal/small-$Z$ part of the
phase diagram around $N=2$.

To understand this regime we consider the itinerant limit. Studying a correlated metallic phase within DMFT translates mainly in characterizing
the coherence scale of the effective impurity problem (self-consistent
atom).  Here we focus on how it is affected by the Hund's coupling. 
The key distinction between different cases is the degeneracy of the
atomic ground-state which is, except for a single electron or hole,
reduced by J, as the state with aligned angular momenta is selected (Table
S1 in the Supporting online information\cite{Janus_Supp}).  Lower degeneracy enhancing the
quantum fluctuations and weaker tunneling from/onto the composite
object suppress the coherence scale\cite{okada73, kusunose97,
  yanase97,yotsuhashi01,nevidomskiy09}, which corresponds to the Kondo scale of the
effective impurity model below which the atomic multiplet is screened
out. Further numerical and analytical results are given in
the Supporting online information\cite{Janus_Supp}.

Now we examine how the different influences of the Hund's rule
coupling are manifested in the physical properties of transition-metal
oxides.  First, a word of warning. Often these materials do not have
perfect cubic symmetry.  The distortions mix the orbitals, reduce the
bandwidths and induce crystal fields which lower the atomic degeneracy
further.  These effects usually enhance the correlations and promote
insulating behavior, which can often be described by an effective
model with a smaller number of orbitals (as some are emptied or filled
by the crystal
fields)~\cite{manini_orbital_2002_prb,pavarini04,poteryaev_crystalfield_prb_2008}.
For illustrative purposes, we choose a set of materials indicated on
Fig.~\ref{fig:Z_t2g_contour}.  In \cite{Janus_Supp}, the
discussion is extended to other materials for which the lifting of
orbital degeneracies is stronger.

We begin with oxides of $3d$ transition metals with a half-filled \t2g
shell, such as SrMnO$_3$ and LaCrO$_3$.  A typical ratio of the
Coulomb repulsion to half bandwidth for these materials is $U/D\simeq
4\mathrm{eV}/1\mathrm{eV}$.  This exceeds substantially the insulating
limit for this case, explaining why no metallic $3t_{2g}^3$ oxides are
known\cite{imada_mit_review, Torrance_why}.  Conversely, at a
comparable value of $U/D$, the $3t_{2g}^1$ cubic SrVO$_3$ is a
moderately correlated metal with $Z^{-1}=m^*/m\simeq
2$\cite{imada_mit_review}.  LDA+DMFT explicitely demonstrates (see
\cite{Janus_Supp}), that SrVO$_3$ would be significantly more
correlated\cite{werner09} were it not for the decorrelating action of
Hund's rule at this filling.  For $3t_{2g}^2$ materials, still within
the same range of $U/D$ (Fig.~\ref{fig:Z_t2g_contour}), strongly
correlated bad-metal behavior caused by the Janus-faced action of $J$
is found.  Observable signatures of bad-metals are large values (beyond the Mott limit~\cite{emery_kivelson_badmetals}) of the
non-$T^2$ but metallic-like resistivity in the extended temperature
range above a very low $T^*$ and a large, poorly screened moment, prone
to itinerant magnetism.  A possible realization among $3d$ oxides is
SrCrO$_3$\cite{chamberland67,zhou06}.

Oxides of $4d$ transition metals are characterized by smaller values of 
$U/D\simeq 2$,  due to the larger bandwidths and smaller screened interaction 
associated with the more extended 4d orbitals. 
We consider the series SrMO$_3$ and Sr$_2$MO$_4$ with 
M = {Mo, Tc, Ru, Rh} (Fig.\ref{fig:Z_t2g_contour}). 
The Technetium compounds are special among those: they are located
very close to the metal/insulator transition.  We are not aware of
transport measurements on these compounds, but a recent
study\cite{rodriguez11} indeed reports antiferromagnetism with a very
large N\'eel temperature $T_N\simeq 1000$~K for SrTcO$_3$.  
Simple model considerations do suggest that the vicinity of the 
Mott critical coupling leads to largest values of $T_N$. 
As a test of our classification, we predict that Sr$_2$TcO$_4$ 
is an insulator or
a very strongly correlated metal.

The Mo-, Ru- and Rh- based compounds are metallic.  For tetragonal
214's an orbital average of the measured values yields
$Z^{-1}=m^*/m\sim 2$ for Sr$_2$MoO$_4$
($4t_{2g}^2$)~\cite{ikeda_Sr2MoO4_jpsj_2000}, $\sim 4$ for
Sr$_2$RuO$_4$ ($4t_{2g}^4$)~\cite{imada_mit_review} and $\sim 2$ for
Sr$_2$RhO$_4$ ($4t_{2g}^5$)~\cite{perry_Sr2RhO4_njp_2006}.  These
variations  are explained by a closer examination of the electronic structure of these
materials. For example, values for Sr$_2$MoO$_4$ and Sr$_2$RuO$_4$
differ because the $t_{2g}$ density of states is not particle-hole
symmetric: the ruthenate has the Fermi level close to a van Hove
singularity and therefore a smaller effective
bandwidth\cite{mravlje11}.  On the other hand, relatively large
correlations found in rhodates occur due to the bandwidth narrowing
and the orbital polarization induced by rotations of the octahedra.
In the regime of weak to moderate correlations with 2 electrons
(Fig. \ref{fig:Z_vs_U}), $Z$ has a steep dependence on $U/D$: this
explains that SrMoO$_3$ has an unusually large metallic conductivity
among oxides\cite{nagai05}.  4d materials can be driven also to the
extreme bad metal regime by rotation-induced narrowing of the
bandwidths achieved by Ca substitution.  An example is Sr$_{1-x}$Ca$_x$RuO$_3$,
which at $x=1$ has $m^*/m>5$ and remains incoherent down to lowest temperatures
measured~\cite{cao97}.



There is thus a class of Hund's correlated materials which are
strongly correlated but driven by $J$ rather than the proximity to a
Mott insulator.  In this respect, we note that the importance of the
Hund's rule coupling has also been emphasized for the iron-based
superconductors\cite{haule09,mazin_Hund_pnictides,aichhorn10}.  With
$6$ electrons in $5$ active orbitals, the bad-metal behaviour observed
for these materials can be attributed to the conflicting action of the
Hund's rule coupling.  This puts pnictides on the map of Hund's
correlated material along with SrCrO$_3$ and SrRuO$_3$ but also raises
important questions. What is the nature of such materials above the
coherence scale and how do they differ from materials close to the
Mott transition?  How are the instabilities to magnetism and
superconductivity affected?

\acknowledgements{
We  are grateful to S.~Biermann, M.~Capone, A.~Fujimori, E.~Gull, K.~Haule,
M.~Imada, A.~Kapitulnik, G.~Kotliar, C.~Martins, I.~Mazin, A.~J.~Millis, Y.~Tokura,
D.~van der Marel, L.~Vaugier and P.~Werner for useful discussions. 
A.G. acknowledges the hospitality of the Universit\'e de Gen\`eve (DPMC) and the support of MANEP, 
and J.M. that of Rutgers University. 
This work was supported by the Partner University Fund, the Agence Nationale de 
la Recherche (ANR-09-RPDOC-019-01 and ANR-2010-BLAN-040804), the RTRA Triangle de la Physique and the Slovenian Research Agency (under contract J1-0747). 
Computer time was provided by IDRIS/GENCI under Grant 2011091393. 
}

\bibliography{Bib/hund,Bib/Janus}

\newpage
\mbox{}

\renewcommand{\thepage}{S\arabic{page}}  
\renewcommand{\thesection}{S\arabic{section}}   
\renewcommand{\thetable}{S\arabic{table}}   
\renewcommand{\thefigure}{S\arabic{figure}}
\newpage
{\bf\Large Supplementary Information}

\section{Model Calculations}
\setcounter{page}{1}
\setcounter{figure}{0}
\setcounter{table}{0}

The 3-band model that we solve in this work can be formulated in term of a tight-binding hamiltonian $H=H_{\rm kin}+H_{\rm int}$ where for the kinetic energy term we take nearest-neighbor hopping amplitudes $t$:
\begin{equation}
H_{\rm{kin}}= \!\! -t\,\sum_m \sum_{<ij>, \sigma}(d^\dagger_{im\sigma}d_{jm\sigma}+\rm{h.c.}).
\end{equation}
Here $d_{im\sigma}^{\dagger}$ creates an electron at site i with spin $\sigma$
in orbital $m$.  Each one of the degenerate bands in this model has a semicircular density of states of half-bandwidth D.

The interaction, in the standard Kanamori form \cite{imada_mit_review} reads
\begin{eqnarray} 
  H_{\rm int}&=&U\sum_m n_{m\uparrow}  n_{m\downarrow} + \\ \nonumber
\!\!\!\!&+&\sum_{m<n,\sigma}\!\![(U-2J) n_{m\sigma} n_{n\bar{\sigma}}  + (U\!\!-3J)n_{m\sigma}n_{n \sigma}]\nonumber \\
    &-& J \sum_{m<n} [ d_{m\uparrow}^{\dagger} d_{m\downarrow} d^{\dagger}_{n\downarrow} d_{n\uparrow} 
    + d_{m\uparrow}^{\dagger} d^{\dagger}_{m\downarrow} d_{n\uparrow} d_{n\downarrow} +
    \rm{h.c.} ]\nonumber \\ \nonumber
   &\overset{t_{2g}}{\mathop{=}}& (U-3J) \frac{\hat{N}(\hat{N}-1)}{2} +  \frac{5}{2} J \hat{N} 
   -2J \vec{S}^2 -\frac{1}{2}J \vec{T}^2.
\end{eqnarray}
The last equality holds only for the \t2g orbitals, but
the rest of the Hamiltonian  remains valid also if
$m$ runs only over \eg orbitals. For \t2g orbitals it differs from
Eq.(1) in the main text for a term $5/2 J \hat{N}$, a simple shift in the chemical potential. 

The model calculations were performed within the
dynamical mean-field theory (DMFT)\cite{georges96}.  In this approach, correlated electron systems are
mapped onto a quantum-impurity problem: an effective atom coupled to a
self-consistent environment. Both the limit of an isolated atom with
its multiplet structure and that of a non-interacting band are
correctly reproduced.

  The associated impurity model was
solved using the zero-temperature exact diagonalization (ED) 
(see e.g. \cite{georges96})
and the continuous-time quantum Monte carlo (CTQMC)
\cite{werner06} methods. 9 bath sites were used in ED calculations,
$10^9$ Monte Carlo cycles per iteration were used in CTQMC calculations at lowest
$kT=D/400$. By comparing results of both methods (see Supplementary Information)
we estimate the typical error in $Z$ as $<10\%$ for $Z\gtrsim0.1$. For
lower $Z$ the coherence scale is smaller than the energy resolution of
the employed methods and the results should be taken as
indicative. We expect the error in $U_c$ to be $<10\%$. The material trends were
verified by LDA+DMFT calculations based on the implementation described in \cite{aichhorn09}. 

\section{Spectrum and degeneracies of the isolated atom}

In Table~\ref{table:eigen_system} we report the spectrum and the
degeneracies of the \t2g  Hamiltonian described by Eq.(1) in the main text.
 Except for the case $N=1$, the degeneracy in every sector is reduced by $J$.

\begin{table}[h]
\begin{ruledtabular}
\begin{tabular}{c c c c c}
N & S & T & Degeneracy $=(2S+1)(2T+1)$ & Energy \\ 
\hline\hline
0 & 0 & 0 & \boxed{1} & 0 \\ \hline
1 & $1/2$ & 1 & \boxed{6} & $-\frac{5}{2} J$  \\ \hline
2 & $1$ & 1 & \boxed{9} & $(U-3J)-5 J$  \\ 
2 & $0$ & 2 & 5 & $(U-3J)-3J $ \\
2 & $0$ & 0 & 1 & $(U-3J) $ \\ \hline
3 & $3/2$ & 0 & \boxed{4} & $3(U-3J)-\frac{15}{2} J $ \\
3 & $1/2$ & 2 & 10 & $3(U-3J)-\frac{9}{2}J$ \\
3 & $1/2$ & 1 & 6 & $3(U-3J)-\frac{5}{2} J$ \\

\end{tabular}
\end{ruledtabular}
\caption{Eigenstates and eigenvalues of the \t2g Hamiltonian in the atomic limit. 
The boxed numbers denote the ground-state degeneracies for $J>0$.  
\label{table:eigen_system}} 
\end{table}


\section{Insights from the impurity model}
The influence of the Hund's coupling on correlations in the impurity
model has been noticed and analysed in several studies,
e.g. \cite{okada73,kusunose97, yanase97, yotsuhashi01, nishikawa10}.
We follow Ref.~\cite{nevidomskiy09} and consider a set of Hund's
coupled Kondo problems defined by the Hamiltonian $H= -J
(\sum_{\alpha=1}^N \vec{S}_\alpha )^2+ J_K \sum_\alpha \vec{S}_\alpha
\cdot \vec{s}^{\,c}_\alpha + \sum_{k \alpha\sigma}\varepsilon_k
c^\dagger_{k\alpha\sigma}c_{k\alpha\sigma}$, with $J_K$ the Kondo
coupling between the spin density of conduction electrons
$\vec{s}^{\,c}_\alpha$ to the local spins $\vec{S}_\alpha$.
In the limit of vanishing Hund's coupling, one
has N uncoupled impurity problems with the Kondo temperature
$T_{K0}=D\exp(-1/\rho J_K)$ where $\rho$ is the density of states of the 
conduction electrons. In the opposite limit of large Hund's coupling a
composite object with large spin $S=N/2$ is formed and in this limit
one can write $\vec{S}_\alpha \sim \vec{S}/N$ 
(a relation verified by summing over $\alpha$). One thus faces a 
$N$-channel Kondo problem coupled to a spin $S$ with
a reduced Kondo coupling $J_K/N$. 
As a result, the Kondo scale $T_K=D\exp(-N/(\rho
J_K))=T_{K0}(T_{K0}/D)^{N-1}$ is reduced and the correlations are
enhanced. 

\begin{figure}
 \begin{center}
   \includegraphics[width=8cm]{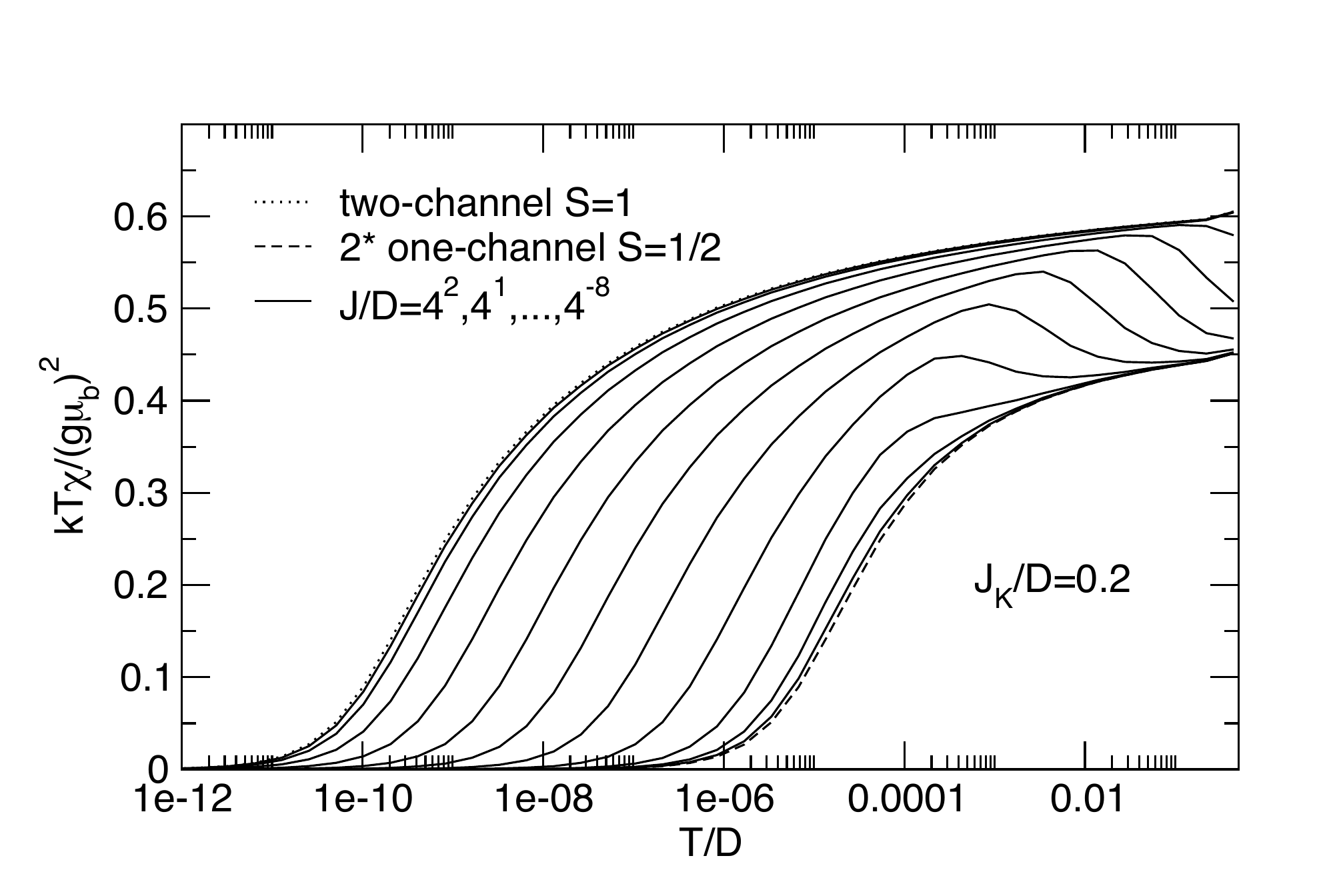}
    \end{center}
 \caption{Local moment $kT\chi /(g \mu_B)^2$   as a
   function of temperature $T/D$. As the Hund's
   coupling $J$ increases the characteristic temperature drops.\label{fig:nrg} }
 \end{figure}

This simple analysis is confirmed by numerical renormalization-group
calculations.  In Fig.~\ref{fig:nrg} we plot the impurity contribution
to the susceptibility for two Hund's rule coupled Kondo impurities
with $S=1/2$ described by the Hamiltonian
\begin{equation}
H= -J (\sum_{\alpha=1,2}
\vec{S}_\alpha )^2+ J_K \sum_\alpha \vec{S}_\alpha \cdot
\vec{s}^{\,c}_\alpha + \sum_{k \alpha\sigma}\varepsilon_k
c^\dagger_{k\alpha\sigma}c_{k\alpha\sigma},
\end{equation}
with $J_K$ the Kondo coupling between the spin density of conduction
electrons $\vec{s}^{\,c}_\alpha$ to the local spins $\vec{S}_\alpha$.
A flat density of states is taken for the conduction electrons. The
Hamiltonian is solved using the numerical-renormalization group
method \cite{zitko}. 

Note that with increasing $J$ the low-energy characteristic scale
below which the impurity is screened (i.e. the Kondo temperature)
diminishes linearly for $J$ larger than the Kondo temperature of the
$J=0$ problem but smaller than $J$ of the order of bandwidth, in
agreement with the perturbative-RG analysis presented in
\cite{yanase97,nevidomskiy09}. Large-$J$ results are interpreted as
discussed in the main text.

\section{Comparison between ED and CTQMC} 
 
\begin{figure}
 \begin{center}
  \includegraphics[width=8cm]{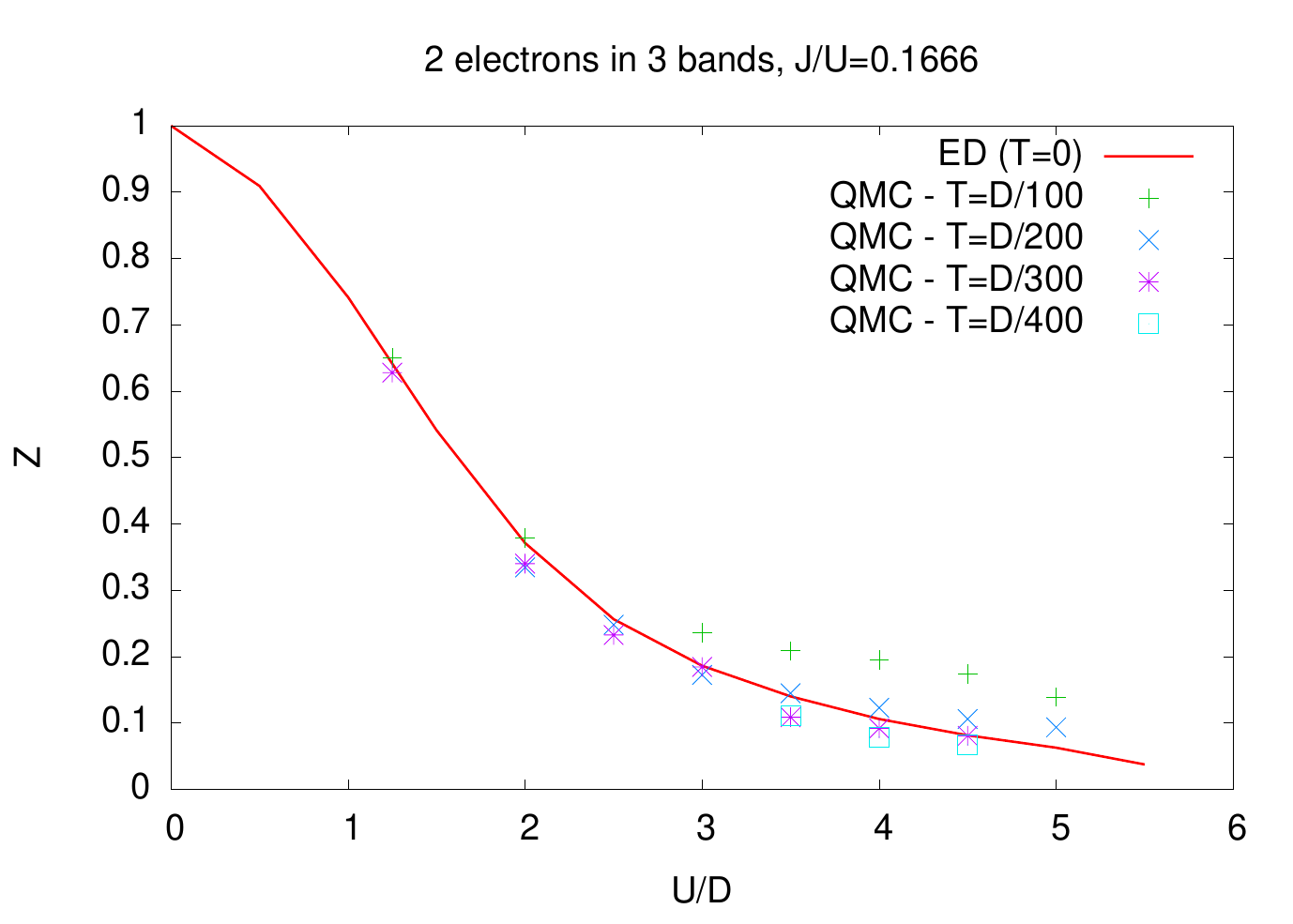}
    \end{center}
 \caption{ Comparison between ED and CTQMC data. \label{fig:methods}}
 \end{figure}

The quasi-particle spectral weights $Z$ are extracted from ED
\cite{caffarel94} and CTQMC \cite{werner06} simulations using
polynomial fits of the electron self-energies for the lowest Matsubara
frequencies. ED is a zero-temperature method (although
finite-temperature-like effects appear due to the discretization of
the bath), therefore a linear approximation gives good results. In
contrast, in CTQMC, the (zero-temperature) $Z$ is estimated from data
at higher temperatures and the polynomial fits use few Matsubara
frequencies (we typically use 4th order polynomial fit to 6 Matsubara
frequencies). 
$Z$ converges only when the temperature is lowered below
the coherence scale\cite{jarrel94}. The convergence of $Z$ as well as
the difference between the ED and CTQMC results are presented in
Fig.~\ref{fig:methods}.  Note that in this data, representative of the
Janus-faced regime, the coherence temperature diminishes rapidly with
$U$ and reaches values we cannot reach numericaly already for $U\sim
4D$, still far from the $U_c\sim 6D$.

Small values of $Z$ indicate also a very low coherence scale $T^*$.

For concreteness, we define $T^*$ by $\Gamma=kT^*$ and
thus identify it with the temperature above which the scattering rate
$\Gamma$  (that grows with temperature) exceeds the typical energy (= temperature) of a quasiparticle.
\cite{mravlje11}.  

For the case of 2 electrons in 3 bands and J/U=0.15 we find $T^*\sim50$K for $Z\sim0.1$; $T^*\lesssim 20$K
for $Z \lesssim 0.05$.  The scattering rates in the $T>T^*$ regime reach large
values and increase monotonously with temperature, a characteristics
of the bad-metals \cite{emery_kivelson_badmetals}. While the low
temperature Fermi liquid behavior is similar for all the fillings, we
find, as in Ref.~\cite{werner08}, a different response at higher temperatures
(or higher frequencies) away from single- and half-filling.

\section{Two-band model}
\begin{figure*}
 \begin{center}
   \includegraphics[width=15cm]{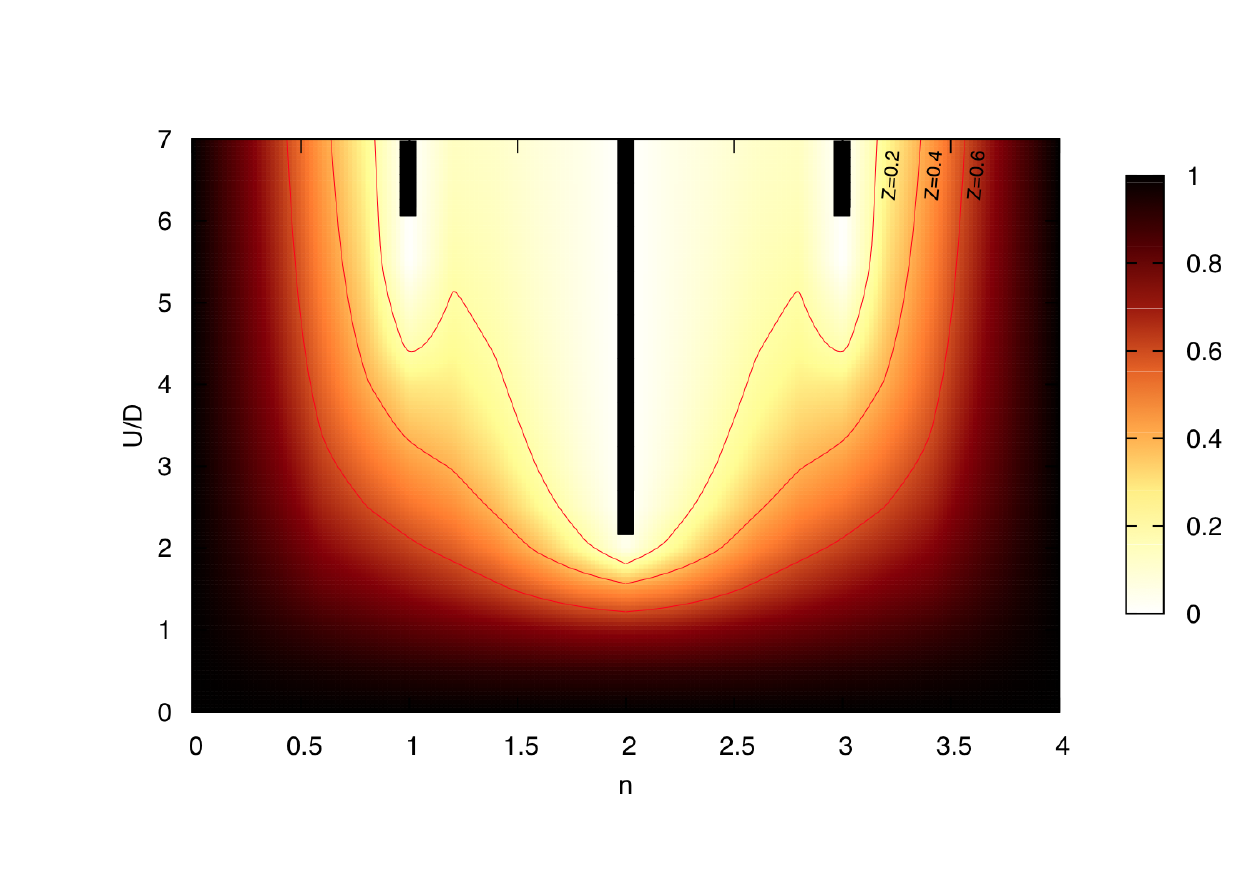}
 \end{center}
 
\caption{Quasiparticle weight $Z$ for a model with 2 orbitals, for
  $J/U=0.15$, as a function of the interaction strength $U$ and
  filling - from empty (0) to full (4). Darker regions correspond to
  good metals and lighter regions to bad metals. The black bars signal
  the Mott insulating phase.\label{fig:Z_eg_contour}}
 \end{figure*}

We solved also the two-band Hamiltonian, relevant to materials with
complete $t_{2g}$ and partially filled $e_g$ shells as well as to
other materials, for instance $4t_{2g}^5$ Sr$_2$RhO$_4$ in which due to the
rotations the $xy-$orbital is pushed below the Fermi level \cite{baumberger06,kim06}.  
In Fig.~\ref{fig:Z_eg_contour} we display the two-band analogue of the data
presented in Fig.~2 of the main text.

The Mott transition occurs for the singly-occupied and the half-filled
case as discussed for the corresponding cases in the three-orbital
model and presented on panels a and c of Fig.~1 in the main text.  In
the singly-occupied case the Hund's coupling has no effect on the
ground state degeneracy but reduces the Mott gap and promotes
metallicity.  On the contrary, in the half-filled case the gap is
increased. Even in the absence of the increase of the gap the
coherence scale would drop as $S=1$ is formed (due to reduced
degeneracy and suppressed tunneling).  Both effects reduce the
coherence temperature, enhance correlations and promote the Mott
insulating state. There is no integer filling at which $J$ has antagonistic effects and bad metals are found only in close proximity to the Mott
insulator.

Note also that, except near half filling, higher correlations
signalled by lower values of $Z$ are found, compared to the three-band
case. This can be related to the lower orbital degeneracy.
Correspondingly, the $U_c$ for the transition to the Mott
insulator is also smaller for the single electron/single hole case,
while for half-filling it is instead slightly larger (for this value
of $J/U=0.15$).  This happens because the Mott gap at half-filling in the
three-orbital model increases as $2J$ whereas in the present case
only as $J$ (see main text).

\section{LDA+DMFT simulations}

To verify that the effects of the Hund's coupling can account for
different behavior of materials, which are chemically and structurally
close, we performed LDA+DMFT simulations within the framework
described in \cite{aichhorn09} on a set of materials noted on Fig.~2 of
the main text. We constructed the t$_{2g}$  Wannier-like orbitals
orbitals from the energy window containing the $t_{2g}$ bands. 

SrVO$_3$ is found to be a moderately correlated material -- as seen
also in experiments \cite{imada_mit_review} -- precisely due to the
decorrelating influence of the Hund's rule coupling in the singly
occupied regime. For $U=4$ we found that by diminishing $J$ from the
realistic value 0.7eV to 0, $Z^{-1}=m*/m$ increases from 2 to 3. At
$U=5$eV and $J=0$ a Mott insulator is found, while for realistic
$J=0.7$eV the material remains a moderately correlated metal with
$m*/m=3$.

The $4t_{2g}^2$ oxide Sr$_2$MoO$_4$ is an electronic analogue of
$4t_{2g}^4$ Sr$_2$RuO$_4$ but found in experiments far more coherent
\cite{ikeda_Sr2MoO4_jpsj_2000}. For Sr$_2$MoO$_4$ we used same
interaction parameters $U=2.3$eV and $J=0.4$eV as for Sr$_2$RuO$_4$
\cite{mravlje11}. Despite the particle-hole symmetry relating the
electron contents of the two compounds, Sr$_2$MoO$_4$ has $m^*/m \sim
2$, twice smaller than $m^*/m \sim 4$ for the Ru-compound. The
distinction occurs because the band structure are not particle-hole
symmetric and the band-value of the density of states close to the
Fermi level is for Sr$_2$MoO$_4$ much smaller, making it less sensible
to the influence of interaction. The SrMoO$_3$ is due to cubic
symmetry and related wider bands  found even more
coherent.  The data on other materials will be reported separately.

\section{Remarks on crystal-field, spin-orbit and charge-transfer effects}
Here we discuss other important aspects determining how
correlated a certain material is, which are only implicitly (through
selecting the effective degeneracy and the effective bandwidth)
contained in simplified model Hamiltonians, e.g. Eq.~(1) of the
main text. 

Lowering of the crystal symmetry from cubic induces the splitting of the $t_{2g}$ orbitals, leading to orbital polarization, diminishes the bandwidths and introduces
significant orbital mixing. The correlations are often enhanced and
especially for $3d$ oxides the distorted materials are
insulating~\cite{manini_orbital_2002_prb,pavarini04,poteryaev_crystalfield_prb_2008,werner07_hilowspin}.

The effect of structural distortions and resulting lowering of
degeneracy are crucial in order to account for the insulating nature
of $3t_{2g}^1$ oxides such as LaTiO$_3$ or YTiO$_3$ \cite{pavarini04}
 and $3t_{2g}^2$ oxides such as LaVO$_3$ or
YVO$_3$~\cite{ray_d2vanadates_prl_2007}.

 For $4t_{2g}^2$ materials the insulating state is induced only
 exceptionally, e.g. in Ca$_2$RuO$_4$ \cite{cao97} 
 and we believe that the other cases in which the structure is distorted are reasonably well described within
 Eq~(1) but with reduced bandwidths, leading to the pronounced bad-metal
 behavior found in these materials\cite{nakatsuji03}.

An analogous effect can appear due to the spin-orbit coupling,
relevant for the late 4d and 5d transition metal elements. The orbital
polarization and the related lower number of active orbitals might
occur also here, thus enhancing correlations (C. Martins and S. Biermann, private communication).

For a large part of the article we have assumed that crystal-field effects are large enough to separate the $e_g$ and $t_{2g}$ manifolds.
On the contrary when the crystal-field effects are small compared to the Hund's coupling,  orbital polarization is suppressed and all five d-orbitals are partially filled as it happens in the case of Iron superconductors cited in the main text. 
In such situations, albeit generalized to a higher number of active orbitals, the effects of Hund's coupling outlined in this article should hold, leading to the filling-dependent promotion of metallic (for single electron/single hole), Mott insulating (for the half-filled case) or bad metallic behaviours (for all other fillings).

In other cases, such as for Manganites, a small crystal field and a large Hund coupling can promote localized $t_{2g}$ electrons in high-spin states and double exchange physics.

Finally, some care in finding a correct low energy Hamiltonian and applying our
arguments must be taken in describing materials which
are of the charge-transfer type, e.g. oxides of the late 3d transition
metals such as SrCoO$_3$, in which the Mott gap is defined by
excitation of holes in the oxygen bands.  The influence of $J$ in tuning the Mott gap in these cases is limited by the presence of these bands, while the suppression of the coherence temperature still applies for
non-singly-occupied cases.

\end{document}